\documentclass[usenatbib,usedcolum,usegraphicx]{mn2e}

\title[Locating lens objects using EAGLE events]{
Constraining the Location of Microlensing Objects \\ towards the LMC
through Parallax Measurement in EAGLE Observations\\ }

\author[T. ~Sumi and Y. ~Kan-ya]
{T. ~Sumi,$^1$ and Y. ~Kan-ya,$^2$ \\
$^1$Princeton University Observatory, Princeton, NJ 08544-1001, USA; 
    e-mail; sumi@astro.princeton.edu\\
$^2$ Division of Earth Rotation, National Astronomical Observatory of Japan, Tokyo 181-8588, Japan; e-mail:yukitoshi.kan-ya@nao.ac.jp }
\date{Accepted 
      Received
      in original form }

\pagerange{\pageref{firstpage}--\pageref{lastpage}}

\begin{document}
\maketitle
\label{firstpage}

\begin{abstract}
  We investigate the possibility of determining whether microlensing
  objects towards the Large Magellanic Cloud (LMC) are in a 
  Galactic thick disc, or are in a Galactic halo, by using parallax 
  measurements with an Earth-radius scale baseline.  
  Our method makes use of EAGLE (Extremely Amplified Gravitational 
  LEnsing) events which are microlensing events with an 
  invisible faint source.  We show that the rate of EAGLE events 
  is as high as that of 
  normal microlensing events, even if they are caused by dark stars 
  in the Galactic thick disc.  We explore the possibility of measuring 
  the parallax effect in EAGLE events towards the LMC by using 
  the {\it Hubble Space Telescope} (HST) or 
  the {\it Very Large Telescope} (VLT).
  We find that EAGLE events enlarge the opportunity of parallax 
  measurements by $4 \sim 10$ times relative to that in normal 
  microlensing events.
  We show that the parallax effect can be measured in $\sim75\%$ (from 
  the HST) and $\sim 60\%$ (from the VLT) of all EAGLE events if most 
  lenses are stars in the Galactic thick or thin disc, while $\sim 20\%$ 
  (from the HST) and $\sim 10\%$ (from the VLT) can be measured if 
  most lenses are halo MACHOs. In combination with the finite source 
  size effect observations, we can strongly
  constrain the location of lenses.  

\end{abstract}


\section{Introduction}

Several groups have carried out gravitational microlensing
observations towards the Large Magellanic Cloud (LMC)
in order to search for MAssive Compact Halo Objects (MACHOs) 
in the Galactic halo.  Until now, 13-17 candidates have been 
found towards the LMC and the microlensing optical depth $\tau$ 
from the events is $1.2^{+0.4}_{-0.3}\times10^{-7}$ (\citealt{alc00b}).
The estimated typical lens mass depends on the
adopted Galactic kinematic model ranging over 
$0.01-1\,M_{\odot}$ (\citealt{alc00b}; \citealt{hon98}).

We have learned that
there are lens objects along the line of sight towards the LMC.
However, the issues of where lens objects are and what they are, are  
still unclear. This is because a degeneracy occurs in ordinary
microlensing events for which  
the amplification is described by (\citealt{pac86})
\begin{equation}
  \label{eq:amp-u}
  A(u)= \frac{u^2+2}{u\sqrt{u^2+4}}\sim\frac{1}{u} \mbox{ for $u\ll1$,} 
\end{equation}
where $u$ is the projected separation of the source and lens in units of 
the the Einstein radius $R_{\rm E}$, which is given by 

\begin{equation}
  R_{\rm E}(M,x) = \sqrt{\frac{4GM}{c^2}D_{\rm s}x(1-x)}.
  \label{eq:re}
\end{equation}
Here $M$ is the lens mass, $x=D_{\rm d}/D_{\rm s}$ is the normalized
lens distance and $D_{\rm d}$ and $D_{\rm s}$ are the observer-lens and
the observer-source star distances. $D_{\rm s}$ is hereafter assumed to
be $50$ kpc. 
The time variation of the parameter $u=u(t)$ is
\begin{equation}
  \label{eq:u}
  u(t)=\sqrt{\beta^{2} + \left( \frac{t-t_{0}}{t_{\rm E}} \right)^2},
\end{equation}
where $\beta$, $t_{0}$, $t_{\rm E}= R_{\rm E}(M,x)/v_{\rm t}$ and
$v_{\rm t}$ are the minimum impact parameter in units of $R_{\rm E}$, 
the time of maximum magnification, the event time-scale and the
transverse velocity of the lens relative to the line of sight 
towards the source star, respectively. From a light curve, one can
determine the value of $\beta$, $t_{0}$ and $t_{\rm E}$, where 
$M$, $x$ and $v_{\rm t}$ are degenerate in $t_{\rm E}$.
This three-fold degeneracy is the essential difficulty in determining the
nature of the lens objects.  There are only marginally possible candidates, 
viz. old white dwarfs (\citealt{alc97,alc00b};
\citealt{han98}), old brown dwarfs (\citealt{hon98}) and primordial
black holes (\citealt{iok00}).

Possibilities for non-halo lensing objects have also been discussed, 
for example, dark objects in a dark heavy component of the LMC itself 
(\citealt{aub99}; \citealt{gyu00}; \citealt{alc01}).  
There is also the possibility 
that the lenses are dark objects in the Galactic thick disc.  While the 
microlensing optical depth $\tau$ by known populations of stars in the 
Galactic thin disc and thick disc is of order $10^{-9}$ 
(\citealt{alc00b}), a maximal heavy thick disc, which may be 
surrounded by an extended dark halo composed of particles, is also a 
possible Galactic component as the reservoir of lenses (\citealt{gat95}; 
\citealt{gat98}). Such a thick disc can have $\tau\simeq7\times10^{-8}$ 
(\citealt{gou94a}; \citealt{gmb94}), so the summed optical depth 
including the contribution from the Galactic thin disc ($\sim2\times10^{-8}$)
and the LMC itself ($\sim1\times10^{-8}$, \citealt{sah94}) can be close 
to the observed value.


The three-fold degeneracy can be resolved 
in some kinds of exotic microlensing events, e.g., the binary event 
(\citealt{har95}; \citealt{alb99}; \citealt{alc99a}; \citealt{hon99}; \citealt{afo00}; \citealt{an01}) and
the finite source transit event (\citealt{gou92,gou94a}; \citealt{nem94};
\citealt{wit94}; \citealt{pen97}).  \citet{sum00} pointed out that an 
extensive transit events search would make it possible
to discriminate between the lenses in the Galactic halo and in the LMC.

The third example is an event with parallax effect, which is essentially 
detected through the difference of light curves due to the spatial 
shift of an observer or observers. In such events we could 
determine the ``reduced transverse velocity'' $\tilde{v}=v_{\rm t}/(1-x)$ 
of the lens. Examples of the parallax effects are the following. The 
annual parallax effect due to the Earth's motion around the Sun during
an event leads an asymmetry in the light curve (e.g. \citealt{gou92};
\citealt{gmb94}; \citealt{miy95}; \citealt{alc95};  \citealt{mao99};
\citealt{ben01}; \citealt{bon01}; \citealt{sos01};  \citealt{smi02}; 
\citealt{mao02}). However, such events 
are rare. This parallax effect generally requires $t_{\rm E}>\sim100$ days, 
while $t_{\rm E}\sim40$ days for typical events. 
Another effect is parallax by the positional difference of two well-separated 
observation sites. In this case, we could measure the relative
difference of the peak amplifications and the time at the peak
amplifications between both. By observing an event from both
a solar-orbit satellite and the Earth, we could utilize the parallax
effect in almost every event, while it is difficult from two distant
observatories on the Earth (\citealt{ref66}; \citealt{gou94b,gou95b}; 
\citealt{hol96}). Furthermore,
the parallax could be measured by the positional shift of an observer due to 
the diurnal motion of the Earth, which was first advanced by \cite{an02}, 
and due to the orbital motion of an Earth-orbit space telescope such as the 
{\it Hubble Space Telescope} (HST) (\citealt{hon99})
in binary events.
In ordinary microlensing events this effect is quite small,
but it is 
more efficient around the peak of high magnification events.

For Galactic bulge events, \citet{gou97} discussed the possibility
of detecting the finite source size and parallax effects by using 
two distinct ground-based telescopes in
order to break the degeneracy in EME's (Extreme
Microlensing Events, $A>200$).
With an approximate estimate for EME observations towards
the LMC, he concluded that it is not feasible.  
However, with a more careful estimate, \cite{nak98} showed that it is 
feasible to detect EAGLE (Extremely Amplified Gravitational LEnsing) 
towards the LMC.
EAGLE is similar to the so-called "Pixel 
lensing" events (\citealt{gou96}) but more simply
defined as the events in which the source star is dimmer than observational 
limiting magnitude (ex. $V_{obs} = 21 \sim 22$), and not concerned
whether the source star is resolved or not.  Some EAGLEs would be EME's. 
EAGLE events could be efficiently detected with 
the ``image subtraction method'' or  ''Difference Image 
Analysis (DIA)'' (\citealt{ala98}; \citealt{ala00}; \citealt{alc99b,alc00a}; 
\citealt{woz00}; \citealt{bon01}), which has been recently developed 
and can perform more accurate photometry than DoPHOT and 
Pixel lensing method.
So, we refer the term EAGLE in this paper.

If lens objects are in the thin or thick disc 
(disc events), $R_{\rm E}$ projected onto the observer plane 
from the source star,
\begin{equation}
  \label{eq:redein}
  \tilde{R_{\rm E}}(M,x)\equiv \frac{R_{\rm E}(M,x)}{1-x}
  \propto\sqrt{\frac{x}{1-x}},
\end{equation}
is much smaller than that for halo events. In this case the light curve is
more sensitive to the small displacement of the observer position.
Therefore the fraction of parallax-measurable EAGLE events out of all
EAGLE events for disc events will be much larger than that for halo
events. 
This fraction is useful to discriminate statistically whether
the lens objects are mainly in the thick disc or not.

Here, we estimate the rate of
parallax-measurable EAGLE events towards the LMC.
In \S\,\ref{sec:eagle} we summarize
the basic equations of EAGLE events. In \S\,\ref{sec:model}
EAGLE event rates are estimated. In 
\S\,\ref{sec:parallax_hst} and \ref{sec:parallax_ground} we describe 
the measurement of parallax effect from space
and ground telescopes, respectively. In \S\,\ref{sec:fraction} 
we calculate the fraction of parallax-measurable events. Discussion 
and conclusion are given in \S\,\ref{sec:disc}.

\section{Basic Formulae for EAGLE events}
\label{sec:eagle}

An EAGLE event is a microlensing event in which a source star 
is fainter than the observational limiting magnitude. A highly 
amplified faint source would be detected as a new star.  
\cite{nak98} showed that the EAGLE event rate is
fairly high towards the LMC. However an EAGLE search has not been 
involved in the microlensing surveys based on 
a DoPHOT-type PSF fitting photometry, which only measure
already detected stars.

A new CCD photometry method called DIA, in which an exposure frame is 
directly compared with a reference frame, enables much more accurate
photometry at any place where any star isn't identified on the reference.
This is more powerful for detecting EAGLE events than the DoPHOT.
Many EAGLE events are expected to be found by DIA.

\subsection{Detection threshold}
\label{sec:basic}

A dim invisible source star with $V$-band 
apparent magnitude $V$ must be amplified brighter than EAGLE 
detection threshold $V_{\rm th}$, which is slightly brighter than the
observational limit $V_{\rm obs}$. So, the threshold amplitude is 
written as
\begin{equation}
  A_{\rm T}(V) \equiv 10^{0.4(V-V_{\rm th})}.
\end{equation}
A corresponding threshold impact parameter $u_{\rm T}=u_{\rm T}(A_{\rm T}(V))$
which is the largest impact parameter to be detected as an EAGLE event 
depends on $V$ and $V_{\rm th}$. This can be approximately written as 
(\citealt{nak98}) 
\begin{equation}
  \label{eq:ut}
  u_{\rm T}(V)\simeq10^{-0.4(V-V_{\rm th})}.
\end{equation}

\section{Galactic model and event rate}
\label{sec:model}

Here we give adopted models of the Galaxy and source
stars, and estimate the event rate. Since only the relative event
rate is discussed, common constant factors $C$, $f$, and the
normalization of the luminosity function are not essential.

When we consider microlensing by lens objects in the maximal disc
(`disc events'), we adopt an exponential disc as the mass density
distribution for the Galactic thin and thick disc. Since the line of
sight to the LMC is almost tangential to the azimuthal direction in the
Galaxy, the mass density at the region efficient to microlensing is not
strongly dependent on the Galactocentric radius $r$ but only on the height
from the disc plane $z$, as long as the disc scale height is much smaller
than $r\simeq8.5$ kpc. So we assume the disc density distribution only
depends on $z$ as
\begin{equation}
   \rho(z)=\frac{\Sigma}{2h}\exp\left( - \frac{z}{h} \right),
   \label{eq:disc}
\end{equation}
where $\Sigma$ is the local disc column density and $h$ is the 
scale height of the disc. For the thin and thick disc we set 
$h_{\rm thin}=350$ pc and $h_{\rm thick}=1400$ pc respectively.
A certain acceptable value is $\Sigma_{\rm thin}\simeq 50$ 
$M_{\odot}$\,pc$^{-2}$ (cf. \citealt{kui89}). If we adopt this 
value, due to the maximal disc limit of $\sim 100$ $M_{\odot}$\,pc$^{-2}$ 
estimated from the rotation speed of the Galaxy (\citealt{bin87}),
$\Sigma_{\rm thick}\simeq 50$ $M_{\odot}$\,pc$^{-2}$ (\citealt{gil83}; 
\citealt{gou94a}; \citealt{gmb94}). However, the value of the
local column density for each disc may be different 
(cf. \citealt{kui89}; \citealt{bah92}; \citealt{cre98}; \citealt{hol00}). 
So we take $\Sigma$ as a model
parameter assuming that the combined mass of the thin and thick discs
does not exceed the maximal disc limit.  We change the fraction of each
component within this limit, i.e., $\Sigma_{\rm thin}(\Sigma_{\rm thick})
= 30(70), 50(50)$ and $70(30)$ $M_{\odot}$\,pc$^{-2}$.

For the disc events we assume the power-law mass function $\phi(M)$ of
lens objects defined by equations (9) and (10) in \cite{sum00}.  
The number density of lens objects with the mass between $M$ and
$M+dM$ is given by $n(M,x)dM=f\rho(z(x))\phi(M)dM/M$, where $f$ is the 
mass fraction of the lens objects to the total mass and is assumed to 
be constant. We take the upper limit of mass $M_{\rm u}=50$ $M_{\odot}$ 
and treat the lower limit $M_{\rm l}$ and the power-law index 
$\alpha_{\rm d}$ as model parameters as $M_{\rm l}= 0.1$ or 
$0.01 M_{\odot}$ and $\alpha_{\rm d} = 2.35$ or $5$, i.e., we assume 
both the ordinary (Salpeter IMF) and an extreme model for a darker 
population of stars as a constituent of the maximal disc. 
Our main results do not depend on this index very much 
(see \S\,\ref{sec:fraction}).
For the luminosity function of the source stars in the LMC, $\phi_{\rm L}(V)$,
we follow equations (15) and (16) in \cite{sum00}
with the IMF index $\alpha_{\rm s} = 2.35$, which is consistent with the
luminosity function of the LMC observed by using the 
HST (\citealt{hol97}).

For MACHOs (`halo events'), we
adopt the spherical `standard' halo model given by (\citealt{alc00b})
\begin{equation}
   \rho_{\rm halo}(r) = \rho_{0} \frac{a^2+r_0^2}{a^2+r^2},
   \label{eq:halo}
\end{equation}
where $\rho_{0}=0.0079$ $M_{\sun}$\,pc$^{-3}$ is the local
mass density, $r$ is the Galactocentric radius, $r_0 = 8.5$ kpc
is the Galactocentric distance of the Sun, and $a=5$ kpc is the core
radius.  We adopt delta-function mass functions with
$M=0.1$ and $0.5$ $M_{\odot}$ because
the lens mass function is not well known.

For the observational parameters we assume $V_{\!\rm obs}=21$, which 
is a typical value for current microlensing programs with 
a 1-m class telescope. The events with the source star having 
$V<21$ are normal microlensing events, while ones with $V>21$ are EAGLE 
events. We consider two different values for the threshold magnitude; 
$V_{\rm th}$ = 19 and 20.

The event rates of normal microlensing events ($\Gamma_{\!\rm N}$) and 
EAGLEs ($\Gamma_{\!\rm E}$), which are proportional to $R_{\rm E}$ and 
$u_{\rm T}R_{\rm E}$ respectively, are given by
\begin{equation}
   \Gamma_{\!\rm N} \! =\! C\!\int_0^{1} \! D_s dx    
             \int_{V_{\rm l}}^{V_{\!\rm obs}} \! dV 
             \int_{M_{\rm l}}^{M_{\rm u}} \! dM 
   R_{\rm E} v_{\rm t} \phi_{\rm L}(V) n(M,x),
   \label{eq:gammaN}
\end{equation}
\begin{equation}
   \Gamma_{\!\rm E} \! =\! C\!\int_0^{1} \! D_s dx  
                        \int_{V_{\!\rm obs}}^{V_{\rm u}} \! dV
                        \int_{M_{\rm l}}^{M_{\rm u}} \! dM 
   u_{\rm T}R_{\rm E} v_{\rm t} \phi_{\rm L}(V) n(M,x),
   \label{eq:gammaE}
\end{equation}
where $V_{\rm u}=30$ and $V_{\rm l}=16$ are the upper and lower limit
of the luminosity function. Note that the integration is performed 
$V<V_{\!\rm obs}$ for $\Gamma_{\!\rm N}$ and 
$V>V_{\!\rm obs}$ for $\Gamma_{\!\rm E}$.

We calculate the relative EAGLE event rate
$\Gamma_{\! \rm E}/\Gamma_{\!\rm N}$
for disc events, with the finite source size effect included.  
The results for $V_{\rm th}=$ 19 and
20 are $\Gamma_{\! \rm E}/\Gamma_{\!\rm N} = 0.73$ and 1.83, 
respectively, only weakly dependent on parameters of the disc 
structure $(\Sigma, h)$ and the mass function ($\alpha_{\rm d}$, 
$M_{\rm l}$). This ratio depends on the 
luminosity function of source stars.  
As a result, only in the 
most extreme case that $V_{\rm th}=19$, $\alpha_{\rm d}=5.0$ and 
$M_{\rm l} = 0.01$ $M_{\odot}$, this ratio is slightly decreased due to 
finite source size effects. In the case that $V_{\rm th}=20$ 
and $\alpha_{\rm d}=2.35$, we found that this effect is not 
negligible only when $M_{\rm l}<10^{-4}$.  For halo events, the events 
affected by finite source size effects are only several percent out of all 
EAGLE events (\citealt{sum00}). So hereafter we neglect the finite source 
size effect.

For events in which the source star is $V>25$, the period in which
the source is visible ($V <21$) is less than 2 days. Then the detection 
efficiency for such events should be very low. So we estimated 
$\Gamma_{\!\rm E}/\Gamma_{\!\rm N}$ in the case that the source star is 
$V<25$, and found $\Gamma_{\!\rm E}/\Gamma_{\!\rm N}  = 0.59$ and 1.47 
with $V_{\rm th}=19$ and 20, respectively.  We checked this ratio in the
conservative case of a gentler slope $\alpha_{\rm s} = 2.0$ with 
$V_{\rm th}=20$ and $V<25$, and found it to be 1.15. This is still 
sufficiently high.


In this paper, we assume that microlensing events (normal and EAGLE) 
would be detected by a 1-m class alert telescope and the real-time analyses 
with DIA, and then follow-up
observations would be performed by an Earth-orbit space telescope or a large
ground-based telescope.

\section{Parallax from Space telescope}
\label{sec:parallax_hst}

Here we describe the parallax effect observed with an Earth-orbit 
space telescope such as the HST. 
In the following analysis, we use the coordinate system which 
is used in \cite{hon99}, i.e., the origin is set to be the centre
of the Earth, and $z$-axis is set to be in the direction of the 
source star. The $x$-axis is set to be perpendicular to both of 
the z-axis and the orbital axis of the space telescope. The 
inclination $i$ of the telescope orbit is defined as an angle 
between the $z$-axis and the orbital axis. We also assume that
the space telescope is in a circular orbit with radius
$r_{\rm st}$ and angular velocity $\omega$. The position of 
the telescope projected onto the observer plane (x-y plane) is 
written as

\begin{equation}
  \label{eq:T}
  {\bf T} = [r_{\rm st}\cos (\omega t +\delta), 
             r_{\rm st}\sin (\omega t +\delta)\cos i].
\end{equation}

Here the angle $\delta$ describes the position of the space telescope 
at $t=t_0$.  Observed from the centre of the Earth, the position of 
the lens object on the lens plane is given by 

\begin{equation}
  \label{eq:LE}
  {\bf L_E} = [v_{\rm t}(t-t_0)\cos \theta - b \sin \theta, 
             v_{\rm t}(t-t_0)\sin \theta + b \cos \theta],
\end{equation}
where $\theta$ is the angle between the $x$-axis and the direction of
the transverse velocity of the lens $v_{\rm t}$ and $b$ is the minimum
physical distance between the lens and the source-Earth line of sight.
The position of the lens relative to the space telescope on the lens 
plane is given by

\begin{equation}
  \label{eq:LT}
  {\bf  L_T} = {\bf L_E} - (1-x){\bf T}. 
\end{equation}
Then from the space telescope $u(t)$ is written as

\begin{flushleft}
\[
u'(t)^2=\left\{\frac{1}{t_{\rm E}}(t-t_0)\cos \theta - \beta \sin \theta
             - \epsilon \cos (\omega t +\delta) \right\}^2  \nonumber 
\]
\end{flushleft}

\begin{equation}
  \label{eq:ust}
       + \left\{\frac{1}{t_{\rm E}}(t-t_0)\sin \theta + \beta \cos \theta
              - \epsilon \sin (\omega t +\delta)\cos i \right\}^2,
\end{equation}
where $\epsilon = r_{\rm st}/\tilde{R}_{\rm E}$.
Equation (\ref{eq:ust}) shows that the parallax effect due to the 
telescope motion causes a wavy trajectory of the lens object relative 
to that from the centre of the Earth.  For orbital parameters for the 
space telescope, we assumed $r_{\rm st}=7000$ km, $i=30^\circ$ and 
$P_{\rm orb} = 2\pi/\omega = 97$ min, which are similar to those of 
the HST.

\begin{figure}
\begin{center}
\includegraphics[angle=0,scale=0.45,keepaspectratio]{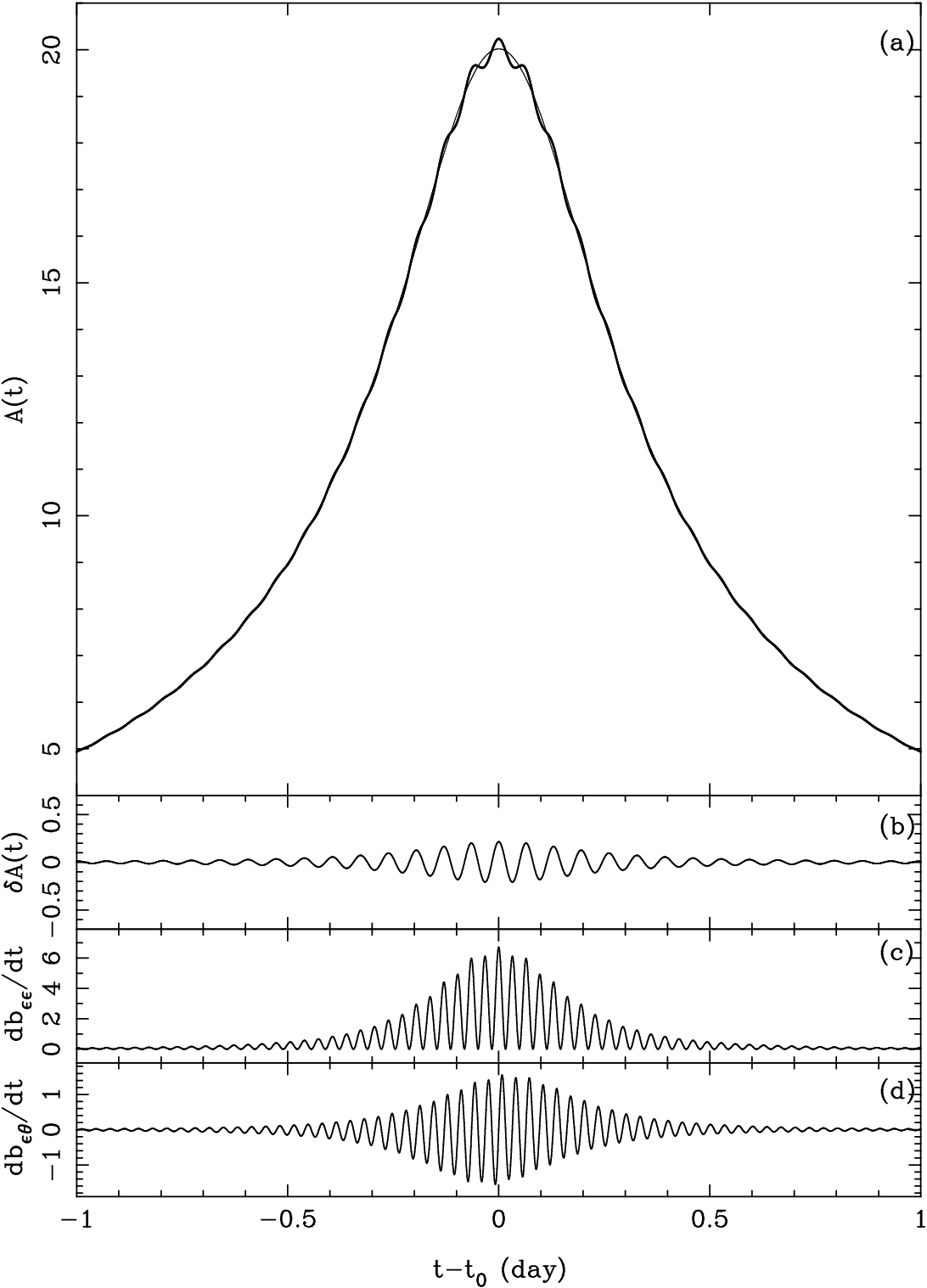}
\caption{(a) Light curves observed from the centre of the Earth 
    (thin line) and from the HST (thick line), (b) $\delta A$ 
    light curve, (c) and (d) show  integrands of the 
    $b_{\epsilon \epsilon}$ and $b_{\epsilon \theta}$, i.e. 
    $\partial b_{\epsilon \epsilon}/\partial t$ in $10^{10}$ e$^-$day$^{-1}$ and 
    $\partial b_{\epsilon \theta}/\partial t$ in $10^{7}$ e$^-$day$^{-1}$ 
    respectively,
    in the case of  $t_{\rm E} = 5$ days, $v_t = 30$ km\,s$^{-1}$, 
    $x=0.01$, $\beta=0.05$, $\theta=90^\circ$ and $\delta=0^\circ$.
    The amplitude of the modulation depends on $\epsilon$. 
}
  \label{fig:lc_hst}
\end{center}
\end{figure}

We show sample light curves without ($A(u)$, thin line) and with
($A(u')$, thick line) the parallax in Fig. \ref{fig:lc_hst}a, 
as well as the difference of the two $\delta A = A(u') - A(u)$ in 
Fig. \ref{fig:lc_hst}b in the case with $t_{\rm E} = 5$ days, 
$v_t = 30$ km\,s$^{-1}$, $x=0.01$, $\beta=0.05$, $\theta=90^\circ$ 
and $\delta=0^\circ$.  Hereafter in the case using the HST, we take 
$\delta=0^\circ$ since $\delta$ is a random parameter related with just 
a phase of orbit. 
We ensured that this does not affect our main results.  In the light 
curve from the space telescope, the small wavy perturbation from that 
observed from the centre of the Earth can be seen.  In the $\delta A$ 
light curve in Fig. \ref{fig:lc_hst}b, we can clearly see that
the fluctuation due to the parallax is enhanced around the peak.

From the observation, we can derive the $\epsilon$ and $\theta$ in 
addition to other parameters $t_{\rm E}$, $\beta$, $t_0$ and $f_0$. 
From $\epsilon$ and $t_{\rm E}$ 
we can derive the ``reduced velocity'' $\tilde{v}$ as follows

\begin{equation}
  \label{eq:tildev}
      \tilde{v} = \frac{r_{\rm st}}{\epsilon t_{\rm E}}.
\end{equation}
The reduced transverse velocity represents the projected relative
transverse velocity between the source star and the lensing object and
will be of great use in investigating the lens location because this
value for each component (halo, thick disc, and thin disc) is different 
(about $240$ km\,s$^{-1}$, $50$ km\,s$^{-1}$ and $30$ km\,s$^{-1}$,
respectively) and hence for disc events $\tilde{v}$ is $\sim 8$ times
smaller than that for halo events.

The rate of parallax-measurable microlensing events
depends on the uncertainty of $\tilde{v}$ which is given by

\begin{eqnarray}
  \label{eq:error_tildev}
    \sigma_{\tilde{v}} &=& 
    \left\{ 
     \left(\sigma_\epsilon\frac{\partial \tilde{v}}{\partial \epsilon}\right)^2
    +\left(\sigma_{t_{\rm E}}\frac{\partial \tilde{v}}{\partial t_{\rm E}}\right)^2
     \right\}^\frac{1}{2} \nonumber  \\ 
    &=& \tilde{v}  \left\{ 
     \left(\frac{\sigma_\epsilon}{\epsilon}\right)^2
    +\left(\frac{\sigma_{t_{\rm E}}}{t_{\rm E}}\right)^2
     \right\}^\frac{1}{2},
\end{eqnarray}
where $\sigma_\epsilon$  and $\sigma_{t_{\rm E}}$ are the errors in the 
$\epsilon$ and $t_{\rm E}$.  We assume that the required threshold 
accuracy to measure $\tilde{v}$ due to the parallax effect is 50\%, 
i.e., $\sigma_{\tilde{v}}/\tilde{v} < 1/2$, in this paper.

The general discussion to estimate the error or the variance of a set of
parameters $a_i$ in fitting a distribution $F(t,a_i)$ is presented by,
e.g., \cite{gou95a,gou98} using the minimum variance bound
given from a well-known theorem in statistics.  The covariance matrix
$c_{ij}$ with respect to $a_i$ for a series of measurements $F(t_{k})$
at time $t_{k}$ with error $\sigma_{k}$ is given by
\begin{equation}
    c_{ij} = b_{ij}^{-1}, \quad  b_{ij} = \sum_{k} \sigma_{F(t_{k})}^{-2}
          \frac{\partial F(t_{k})}{\partial a_{i}}
          \frac{\partial F(t_{k})}{\partial a_{j}}.
\label{eq:cova}
\end{equation}
The variance of $a_{i}$ is just the diagonal elements $c_{ii}$.

We take $F(t)$ as the number of the detected photo-electrons at time
$t$ from the source star. During a short time
interval $T$, we get $F(t)=(f_{0}(V) A(t;\beta, t_{\rm E},t_{0}) +f_b)T$, 
where $f_{0}$ denotes the average photo-electron flux
from the unamplified source star and $f_b$ denotes the background flux 
in the PSF aperture from the sky and the 
unlensed blending stars.  $A(t;\beta, t_{\rm E}, t_{0})$ is the
amplification produced by microlensing given by equations
(\ref{eq:amp-u}) and (\ref{eq:u}).  By using the DIA, the signal is 
expressed as $\Delta F(t)= f_{0}(A-1)$ and $\sigma_F(t)=\sqrt{F(t)}$.
Taking the limit $T \to 0$, $b_{ij}$ in equation (\ref{eq:cova})
is given by
\begin{equation}
  \label{eq:bij}
  b_{ij} \simeq
   \int_{t_{\rm begin}}^{t_{\rm end}}
  \frac{1}{f_{0}(V)A+f_b} \left[\frac{\partial f_{0}(A-1)}{\partial a_i}
  \right] 
                          \left[\frac{\partial f_{0}(A-1)}{\partial a_j}
  \right] dt.
\end{equation}
A parallax effect tends to 
be measured accurately in the case that $f_{0}$ is large, and/or
$\beta$, $v_{\rm t}$, $x$ and/or $M$ are small.

We show the integrands of $b_{\epsilon \epsilon}$ and 
$b_{\epsilon \theta}$ in Fig. \ref{fig:lc_hst}c and 
\ref{fig:lc_hst}d, respectively, in the case of the source 
magnitude of $V=21$ and with the same parameters as 
in Fig. \ref{fig:lc_hst}a. Here we assumed 
$f_0(V=20) = 350$ e$^-$\,s$^{-1}$ and $f_b = 10$ e$^-$\,s$^{-1}$, 
which is similar to those of the HST with the {\it Advanced 
Camera for Surveys} (ACS) (\citealt{pav01}). $f_b$ in the aperture 
of the PSF is estimated by taking account of the following sources;
(1) Zodical light $V_{\rm ZL}= 23.3$ mag\,arcsec$^{-2}$, which is 
the smallest because the  Ecliptic latitude of the LMC is $\sim 90^\circ$.
(2) Earth shine from the limb of the sunlit Earth $V_{\rm ES} = 21.4$ 
mag\,arcsec$^{-2}$, which is the mean when the angle $\theta_{\rm limb}$ 
between the target and the bright Earth limb is larger than 12$^\circ$.
(3) Light from Blending stars $V_{\rm LMC} = 22.01$ mag\,arcsec$^{-2}$, 
which is the mean $V$-band surface brightness of the inner 10 deg$^2$ of 
the LMC (\citealt{vau57}). We reduced $f_0(V)$ and $f_b$ by a factor of 
$3/4$ because the LMC could not be observed when the HST is in shadow 
of the Earth and we limit $\theta_{\rm limb} > 12^\circ$.
(Note: the phase when the observation would be interrupted is not 
always the same, because the actual orbit of the HST is not circle but  
the latitude of the HST is changing between $-30^\circ$ and $30^\circ$.)
We multiplied $f_0(V)$ and $f_b$ by $0.9$ taking account of 
the dead time of camera assuming $\sim5$ minutes exposures.

In Fig. \ref{fig:lc_hst}c and \ref{fig:lc_hst}d, we can see that 
most of the information of $\epsilon$ is in very short period 
($\sim t_{\rm E}\beta$) around the peak, and the correlation between 
$\epsilon$ and $\theta$, i.e., $b_{\epsilon \theta}$ is negligible in 
a day-period observation.  Furthermore, we ensured that $\epsilon$ is 
completely independent of other parameters.

\begin{figure}
\begin{center}
\includegraphics[angle=-90,scale=0.3,keepaspectratio]{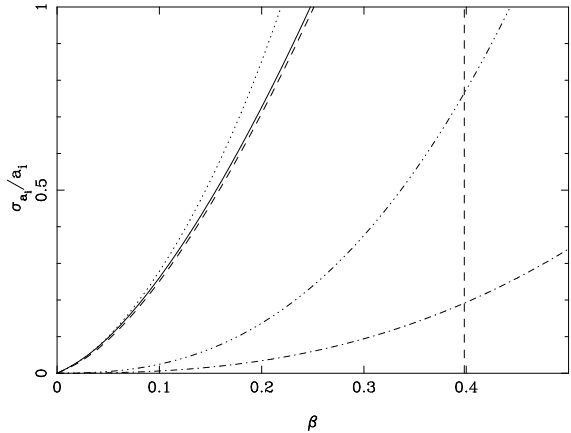}
\caption{Uncertainty $\sigma_{a_i}/a_i$ as a function of the impact 
  parameter $\beta$ for $\epsilon$ (solid line), $\theta$ (dashed line),
  (i) $t_{\rm E}$  (dotted line) and  $t_{\rm E}$  under the condition 
  that (ii) $f_0$ (3dotted-dashed line) or (iii) both $f_0$ and $t_0$ 
  (dot-dashed line) are externally constrained, 
  in the case of $t_{\rm E} = 40$ days, $x= 0.01$, 
  $v_{\rm t}=30$ km\,s$^{-1}$, $\theta=45^\circ$ and the source magnitude 
  $V=21$.
  The vertical dashed line indicates the EAGLE detection threshold 
  impact parameter $u_{\rm T}$.
}
  \label{fig:NS_hst}
\end{center}
\end{figure}

We assume that the follow-up observation for EAGLE events would 
be carried out by the HST for one day from $t_0$. The extensive 
follow-up observations from ground-based small telescopes around 
the world make it possible to know $t_0$.
We evaluate the uncertainty of the parameters  $\sigma_{a_i}$ 
($a_i = \epsilon, \theta, t_{\rm E}$).  In Fig. \ref{fig:NS_hst}, 
we show $\sigma_{a_i}/a_i$ as a function of $\beta$ for the 
case of the typical event time-scale of $t_{\rm E} = 40$ days with 
same parameters used above except for $\theta=45^\circ$. 
The $\sigma_\epsilon$ is the minimum at $\theta=90^\circ$ and the 
maximum at $\theta=0^\circ$. Hereafter, we fixed $\theta=45^\circ$ 
since $\theta$ is a random parameter. We ensured that the difference 
of the main results in \S\,\ref{sec:results} between the case 
using fixed $\theta=45^\circ$ and the case taking $\theta$ at 
random, is negligible.

Meanwhile $t_{\rm E}$ is highly correlated with
other parameters, i.e., $\beta$, $t_0$ and $f_0$.  
If the event is observed symmetrically in time around $t_0$, 
$t_{\rm E}$ is completely independent of $t_0$ because $t_0$ is an odd 
parameter in time.  However, in the case that the event would be observed
for a day from $t_0$, $t_{\rm E}$ is no longer independent of $t_0$. 
In Fig. \ref{fig:NS_hst} we show (i) $\sigma_{t_{\rm E}}/t_{\rm E}$ 
(dotted line) and $\sigma_{t_{\rm E}}/t_{\rm E}$ under the condition 
that (ii) $f_0$ (3dotted-dashed line) or (iii) both $f_0$ and $t_0$ 
(dot-dashed line) are externally constrained.
This (i) $\sigma_{t_{\rm E}}$  in which there is 
a $f_0$ would be much improved over using two different telescopes 
to measure the parallax in which there are two $f_0$.  Though, comparing 
(i) and (ii), the former one deteriorates relative to  the latter one 
because of degeneracies among the parameters $t_E$, $\beta$ and $f_0$, 
which is severe in the case that the event is observed only around the peak 
(\citealt{gou97}; \citealt{han97}). This can be improved by constraining 
$f_0$ by follow-up observation with the HST after the event 
(\citealt{han97}). Hereafter we assume that $f_0$ would be constrained 
by the follow-up observation.
As shown in Fig. \ref{fig:NS_hst}, case (ii) is larger than (iii), 
but still quite small relative to $\sigma_{\epsilon}/\epsilon$.
Especially in the region of $\sigma_{\epsilon}/\epsilon<1/2$, 
where we are concerned, $\sigma_{t_{\rm E}}/t_{\rm E}$ is at 
least 10 times smaller than $\sigma_{\epsilon}/\epsilon$. 
In the case for brighter source event ($V<21$), the contribution of 
$\sigma_{t_{\rm E}}/t_{\rm E}$ becomes slightly larger at 
$\sigma_{\epsilon}/\epsilon \sim 1/2$, but in this case $t_0$ could be
constrained well from the overall light curve taken by ground-based telescopes.
Then hereafter we neglect $\sigma_{t_{\rm E}}/t_{\rm E}$ in equation 
(\ref{eq:error_tildev}), i.e., we can rewrite equation (\ref{eq:error_tildev})
to $\sigma_{\tilde{v}}/\tilde{v} \simeq \sigma_\epsilon/\epsilon$.

From the curve of $\sigma_{\epsilon}/\epsilon$, we can obtain the
critical impact parameter $\beta_{\rm crit}$ to detect an parallax effect 
with 50\% accuracy, which corresponds to $\sigma_{\epsilon}/\epsilon = 1/2$.

\section{Parallax from Ground Telescope}
\label{sec:parallax_ground}

Here we consider parallax measurements from an 8-m class telescope
at a latitude of $-30^\circ$ on the Earth such as the 
{\it Very Large Telescope} (VLT). In this case 
the light curve would be expressed with the same equation as that from the 
HST (see equation (\ref{eq:ust})) except that the radius 
of the orbit is $R_{\rm orb} = R_\oplus \cos30^\circ = 5500$ km, 
the period is $P_{\rm orb} = 2\pi/\omega = 1$ day and we can observe 
events only during night-time.  Here $R_\oplus$ is the Earth radius.

\begin{figure}
\begin{center}
\includegraphics[angle=0,scale=0.45,keepaspectratio]{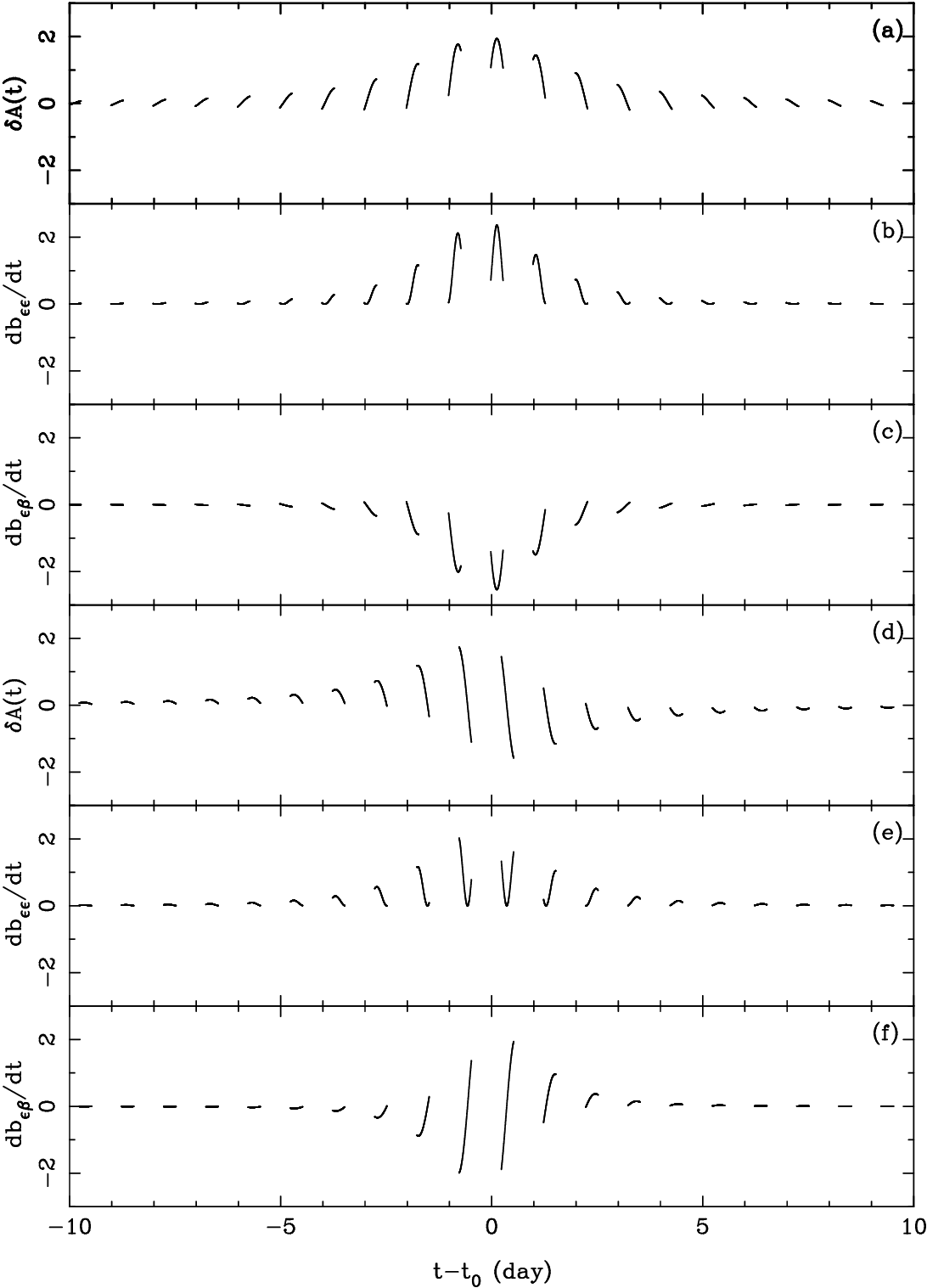}
\caption{ Light curves of $\delta A$ (a and d) in $10^{-2}$ and 
integrands of $b_{\epsilon \epsilon}$ (b and e) and $b_{\epsilon \beta}$ 
(c and f) in $10^{-11}$ in the case of $t_{\rm E} = 40$ days, 
$v_{\rm t} = 30$ km\,s$^{-1}$, $x=0.01$, $\beta=0.05$ and 
the source magnitude $V=21$.  Panels 
(a)-(c) and (d)-(f) represent the cases of $\theta_N - \theta = 90^\circ$, 
and the case of $\theta_N = \theta$  respectively.  
}
  \label{fig:lc_VLT}
\end{center}
\end{figure}

In Fig. \ref{fig:lc_VLT}, we show sample light curves of 
$\delta A = A(u') - A(u)$ (a and d) and integrands of 
$b_{\epsilon \epsilon}$ (b and e) and  $b_{\epsilon \beta}$ (c and f) 
in the case of $t_{\rm E} = 40$ days, $v_{\rm t} = 30$ km\,s$^{-1}$, 
$x=0.01$ and $\beta=0.05$. Here we assumed that 
observations are performed for 7 hours a day.
Fig. \ref{fig:lc_hst}a-c and \ref{fig:lc_hst}d-f represent the case that the 
dark-side of the Earth is near to the lens relative to the centre of 
the Earth, i.e.  $\theta_{\rm N} - \theta = 90^\circ$, and the case that 
$\theta_{\rm N} = \theta$  respectively, where  $\theta_{\rm N}$ is
the angle between the centre of the observation night relative to
the centre of the Earth and the x-axis.  
In this case, we can observe only some parts of the wavy light curve,
and then $\epsilon$ is highly correlated with
other parameters, i.e., $\theta$, $t_{\rm E}$, $\beta$, $t_0$ and $f_0$.
In the case of Fig. \ref{fig:lc_VLT}a-c, $b_{\epsilon \beta}$ becomes
large as well as $b_{\epsilon \epsilon}$, then $\sigma_\epsilon$ becomes
significantly larger than the case of Fig. \ref{fig:lc_VLT}d-f in which 
$b_{\epsilon \beta}$ becomes quite small, as well as the other cross terms 
$b_{ij}$.

We estimate $\sigma_{\epsilon}$ by the same procedure in 
\S\,\ref{sec:parallax_hst}, i.e. evaluating the covariant matrix.
In this case, we assume the photo-electron flux is 
$f_{0}(V=20)\simeq2000$ e$^-$\,s$^{-1}$ using an 8-m class telescope 
and thin CCD cameras with a filter near the V band with a relatively narrow 
bandwidth ($\sim 100$ nm). The very broad-band filter is not advisable
because that might not be able to eliminate the effect of differential
refraction or differential extinction of the atmosphere (\citealt{gou98}).
We reduce $f_{0}(V)$ and $f_b$  by a factor of 2 to take account of
weather conditions and CCD camera dead time.
We adopt the background flux $f_{\rm b} = 1200$ e$^-$\,s$^{-1}$ 
corresponding to $V=20.6$ mag, which is estimated as follows:
the mean V-band surface brightness of the inner 10 deg$^2$ of 
the LMC is $V=22.01$ mag\,arcsec$^{-2}$, and the sky value is 
$V= 21.6$ mag\,arcsec$^{-2}$ (\citealt{vau57}). We assume the 
aperture of the PSF is $0.49 \pi$ arcsec$^2$ 
($\sim 0.7''$ seeing) which is similar to the typical seeing of the VLT. 
The total brightness in the
aperture is 20.6 mag.  We adopt this value for $f_{\rm b}$.
Here we assumed that the follow-up observation would be taken 
for three nights ( 7 hours a night) around the peak.

The LMC is visible enough ($\sim$ 7 hours) only during the southern
summer (between equinoxes) from the site at a latitude of $-30^\circ$, 
which corresponds to that $\theta_{\rm N}$ is distributed at random between 
$0^\circ\sim 180^\circ$.  $\sigma_{\epsilon}$ is maximum at 
$\theta_{\rm N} = 0^\circ$ and $180^\circ$ and minimum at  
$\theta_{\rm N} = 90^\circ$ because of the inclination of 
the orbit axis $i = 30^\circ$.
Hereafter we use the mean value of $\theta_{\rm N} = 45^\circ$.
We ensure that the difference of the main results in the following 
analyses between the case using the fixed $\theta_{\rm N} = 45^\circ$ 
and that using a random $\theta_{\rm N}$ is negligible. 
Anyway, this effect
is much smaller than the dependence on $\theta - \theta_{\rm N}$.

\begin{figure}
\begin{center}
\includegraphics[angle=0,scale=0.4,keepaspectratio]{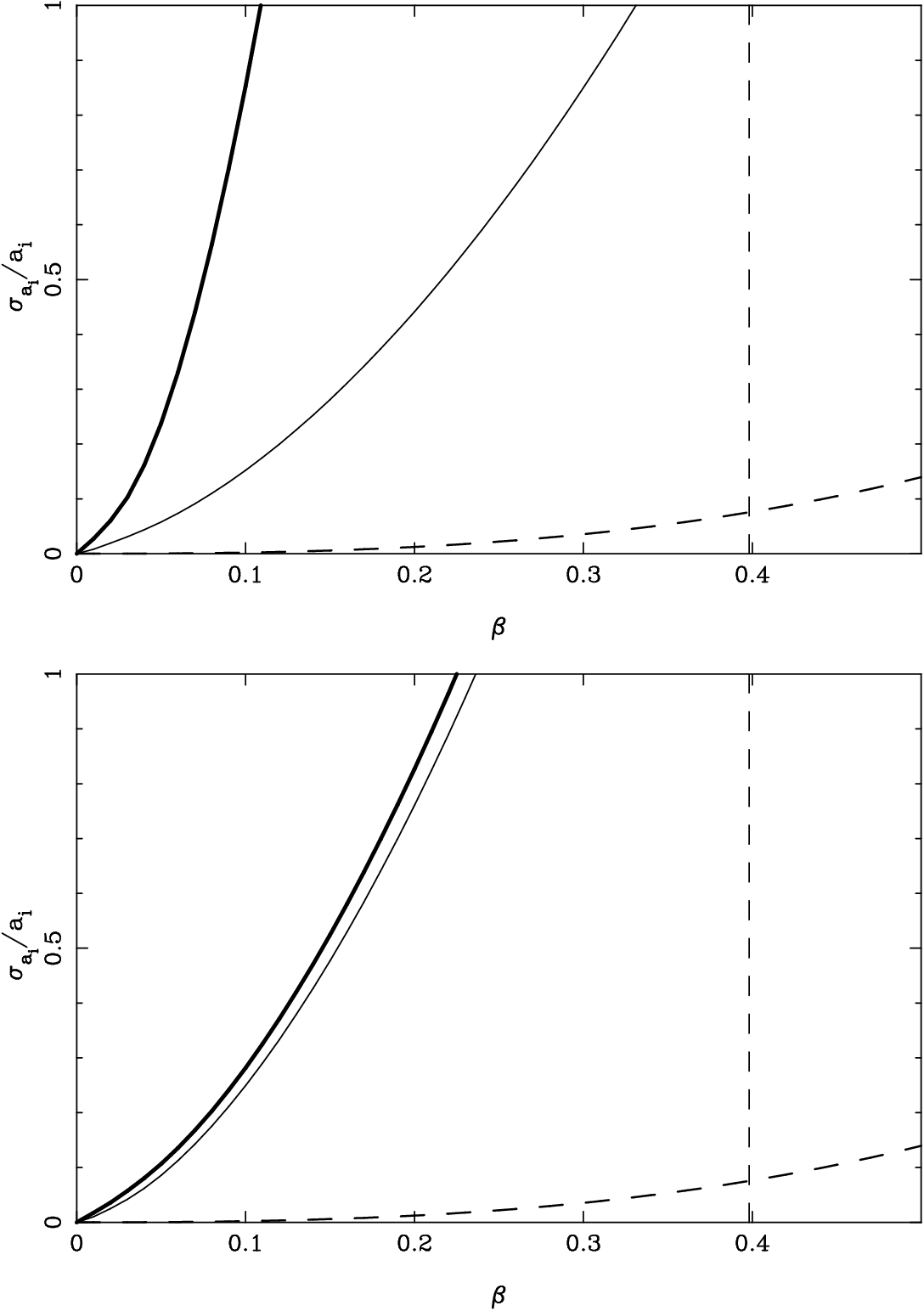}
\caption{Uncertainty $\sigma_{\epsilon}/\epsilon$ under the condition that 
$f_0$ (solid line) and all other parameters (dotted line) are externally 
constrained, and $\sigma_{t_{\rm E}} /t_{\rm E}$  under the condition that 
$f_0$ is externally constrained (dot-dashed line) as a function of 
$\beta$ in the case of $t_E = 40$ days, $x= 0.01$, $v_t=30$ km\,s$^{-1}$, 
$\theta_{\rm N}=45^\circ$, $\delta=45^\circ$ and the source magnitude $V=21$. 
Upper and lower panels represent the case of $\theta=45^\circ$
(i.e. $\theta_{\rm N} = \theta$, which correspond to Fig.
\ref{fig:lc_hst}a-c) and the case of $\theta=135^\circ$ 
(i.e. $\theta_{\rm N} - \theta =  90^\circ$, which correspond to 
Fig. \ref{fig:lc_hst}d-f) respectively.
The vertical dashed line indicates $u_{\rm T}$.
}
  \label{fig:NS_ground}
\end{center}
\end{figure}

In Fig. \ref{fig:NS_ground},
we show $\sigma_{\epsilon}/\epsilon$ under the condition that $f_0$ 
is externally constrained (thick line),  $\sigma_{\epsilon}/\epsilon$
under the condition that all other parameters are externally constrained
(thin line), which is shown to compare with the case using the HST,  and  
$\sigma_{t_{\rm E}}/t_{\rm E}$ under the condition that $f_0$ is 
externally constrained (dashed line) as a function of $\beta$ in the 
case of $t_{\rm E} = 40$ days, $x= 0.01$, $v_{\rm t}=30$ km\,s$^{-1}$,
$\theta_{\rm N}=45^\circ$, $\delta=45^\circ$
and the source magnitude of $V=21$.
Upper and lower panels represent the case of $\theta=-45^\circ$
(i.e. $\theta_{\rm N} - \theta = 90^\circ$, which correspond to Fig. 
\ref{fig:lc_hst}a-c) and the case of $\theta=45^\circ$ 
(i.e. $\theta_{\rm N} = \theta $, which correspond to 
Fig. \ref{fig:lc_hst}d-f), respectively.
In this figure, $\sigma_{\tilde{v}}/\tilde{v}$ heavily
depends on  $\theta_{\rm N} - \theta$,
it especially deteriorates in the case of 
$\theta_{\rm N} - \theta  = 90^\circ$  relative to the thin line.
In this case $\sigma_{t_{\rm E}}/t_{\rm E}$ is also negligible.

\begin{figure}
\begin{center}
\includegraphics[angle=-90,scale=0.3,keepaspectratio]{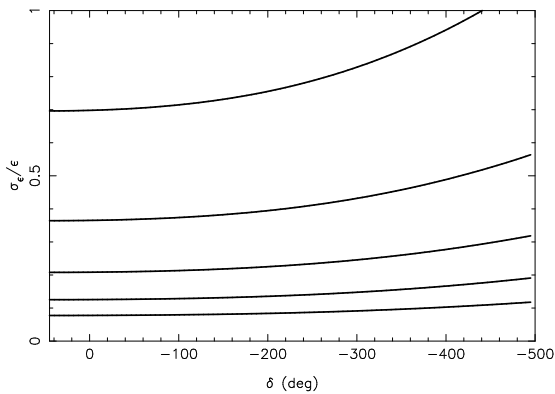}
\caption{Uncertainty $\sigma_{\epsilon}/\epsilon$ as a function of
$\delta$ in the case of using the same parameters as Fig.
\ref{fig:NS_ground} and $\beta = 0.04$,
for the source magnitude of $V= 20, 21, 22, 23$ and $24$ from bottom to top.
Here smaller $\delta$ means that observations are delayed relative to
the optimal case that the peak is observed at the centre of the second 
night ($\delta = 45^\circ$).
}
  \label{fig:NS_delta}
\end{center}
\end{figure}

$\delta=45^\circ$ is the most optimal case, i.e. one observes the 
peak at $\theta_{\rm N}$ in the second observation night.  In actual 
$\delta$ depends on random parameters $t_0$ and $\theta_{\rm N}$, 
and how early one can start the follow-up. Fig. \ref{fig:NS_delta} show
$\sigma_{\epsilon}/\epsilon$ as a function of  $\delta$ in the case
using same parameters in Fig. \ref{fig:NS_ground} and $\beta = 0.04$,
for the source magnitude of $V= 20, 21, 22, 23$ and $24$ (from bottom to top).
Here we take the mean value of $\sigma_{\epsilon}/\epsilon$ over
$-45^\circ<\theta<135^\circ$ because the $\theta$ at which  
$\sigma_{\epsilon}/\epsilon$ is the maximum or 
minimum, is related with $\delta$.
In Fig. \ref{fig:NS_delta} we can see $\sigma_{\epsilon}/\epsilon$
does not depend on $\delta$ very much, as long as observations are spread 
roughly around $t_0$, i.e. it does not need to cover the peak exactly. 
When the follow-up observation is delayed more than 1 day 
($\delta < \sim -300$), it becomes worse especially for dimmer source 
events. This is worse for shorter time-scale events with
much smaller $\beta$ because most of the information of the parameters 
is contained in a very short period ($\sim t_{\rm E}\beta$) around the peak.
Furthermore in extreme cases ($t_{\rm E}\ll 10$ days, $\beta \ll0.01$), 
generally $\sigma_{\epsilon}/\epsilon$ is small, however, it depends on
whether the peak is covered exactly or not. 
This effect is large in dimmer source
events ($V >25$).

From the curve of $\sigma_{\epsilon}/\epsilon$ we can also obtain 
$\beta_{\rm crit}$.
We take the mean of $\beta_{\rm crit}$  over $-45^\circ<\theta<135^\circ$.
Uncertainty $\sigma_{\epsilon}$ also depends on the direction of $v_t$.
Note that this is negligible in the case using the HST.
We take the mean of $\beta_{\rm crit}$ on the cases of $-v_t$ and $+v_t$ 
(i.e. same and opposite direction relative to the Earth rotation, 
respectively) in \S\,\ref{sec:results}.

\section{Fraction of Parallax-measurable events}
\label{sec:fraction}

Here we calculate the fraction of parallax-measurable
events out of all events in the case of $V_{\rm th} = 20$ 
and $V_{\!\rm obs} = 21$.  We adopt the same Galaxy density 
model and the source luminosity function  as \S\,\ref{sec:model}.
To make the estimation realistic,
we applied a photometric error $17\%$ larger than the photon noise,
which is estimated for the DIA (\citealt{woz00}).
Furthermore, we specify the typical (transverse) velocity of the lens objects 
to estimate parallax-measurable event rate. 

\subsection{Model kinematics}
\label{sec:velocity}

For disc events the typical transverse
velocity of a lens object $v_{\rm t}$ is estimated as follows. First
we assume that the thin and thick discs have the same rotation velocity and
the relative velocity between the mean flow of the lens objects and
the local standard of rest (LSR) is zero. In fact the effect of the
drift velocity of the thick disc relative to the thin disc due to the
large velocity dispersion of the thick disc is negligible compared to
other effects.  $v_{\rm t}$ is the composition of the tangential
component of the mean velocity of the Earth relative to the LSR $v_{\oplus}$
and the velocity dispersion of the disc $\sigma$, which is assumed to
be isotropic for both discs, as
\begin{equation}
   v_{\rm t}^{2}(z) = 2\sigma^{2}(z) + v_{\oplus}^{2}.
   \label{eq:vt}
\end{equation}
We take $v_{\oplus}= 30$ km\,s$^{-1}$, which includes the orbital velocity
of Earth around the Sun of $30$ km\,s$^{-1}$ and a velocity of the Sun
relative to LSR of 20km s$^{-1}$ (\citealt{miy98}). $\sigma^{2}(z)$ and
$v_{\rm t}^{2}$ are assumed to be a function of $z$ only (see the
discussion on equation (\ref{eq:disc})).  The density profile of the
disc is same as in \S\,\ref{sec:model}, i.e., the exponential disc density
distributions in equation (\ref{eq:disc}) for thin ($h_{\rm
  thin}=350$ pc) and thick ($h_{\rm thick}=1400$ pc) discs with variable
values of the local disc column density as $\Sigma_{\rm thin}
(\Sigma_{\rm thick}) = 30(70)$, 50(50), 70(30) $M_{\odot}$pc$^{-2}$.
Solving the Poisson equation for the density profile composed of the
thin and the thick discs $\sigma^{2}(z)$ is obtained from the
$z$-component of the Jeans equation (e.g.  \citealt{bin87}) as
\begin{equation}
  \sigma^{2}(z)= \frac{\pi G}{2}
  \frac{ \left[\Sigma_{\rm thin} \exp
   \left( -\frac{\displaystyle z}{\displaystyle h_{\rm thin}}\right)
  + \Sigma_{\rm thick} \exp \left(
   -\frac{\displaystyle z}{\displaystyle h_{\rm thick}}\right)\right]^2}
  { \rho_{\rm thin}(z) + \rho_{\rm thick}(z)}.
  \label{eq:sigmaz}
\end{equation}

For halo events we adopt the same model as in \S\,\ref{sec:model} and
we assume $v_{\rm t} = 190$ km\,s$^{-1}$ or $220$ km\,s$^{-1}$. $v_{\rm t}
= 190$ km\,s$^{-1}$ is more plausible because of the slight contribution
from the disc gravity to the rotation curve, which is assumed to be
flat with $v_{\rm rot}=220$ km\,s$^{-1}$.

\subsection{Results}
\label{sec:results}

We estimate the relative parallax-measurable event rate.
The parallax-measurable event rate 
is written as
\[
\Gamma_{\!\rm P} = C
  \int^{1}_{0}D_{\rm S}dx
  \int^{V_{\rm u}}_{V_{\rm l}}dV
  \int^{M_{\rm u}}_{M_{\rm l}}dM\
\]
\begin{equation}
  \label{eq:betathut}
   \beta_{\rm crit}(V,M,x,v_{\rm t}(z(x)))
  R_{\rm E}v_{\rm t} \phi_{\rm L}(V)
  \,n(M,x),
\end{equation}
where $C$ is the constant common to the expression of
$\Gamma_{\!\rm N}$ (equation (\ref{eq:gammaN})) and $\Gamma_{\!\rm E}$ 
(equation (\ref{eq:gammaE})). And $V_{\rm u} = V_{\!\rm obs}$ for the 
normal events and $V_{\rm l} = V_{\!\rm obs}$ for EAGLEs.
$\beta_{\rm crit}$ is estimated by same procedure presented in
\S\,\ref{sec:parallax_hst} and \ref{sec:parallax_ground} for each 
parameter.

\begin{figure}
\begin{center}
\includegraphics[angle=-90,scale=0.35,keepaspectratio]{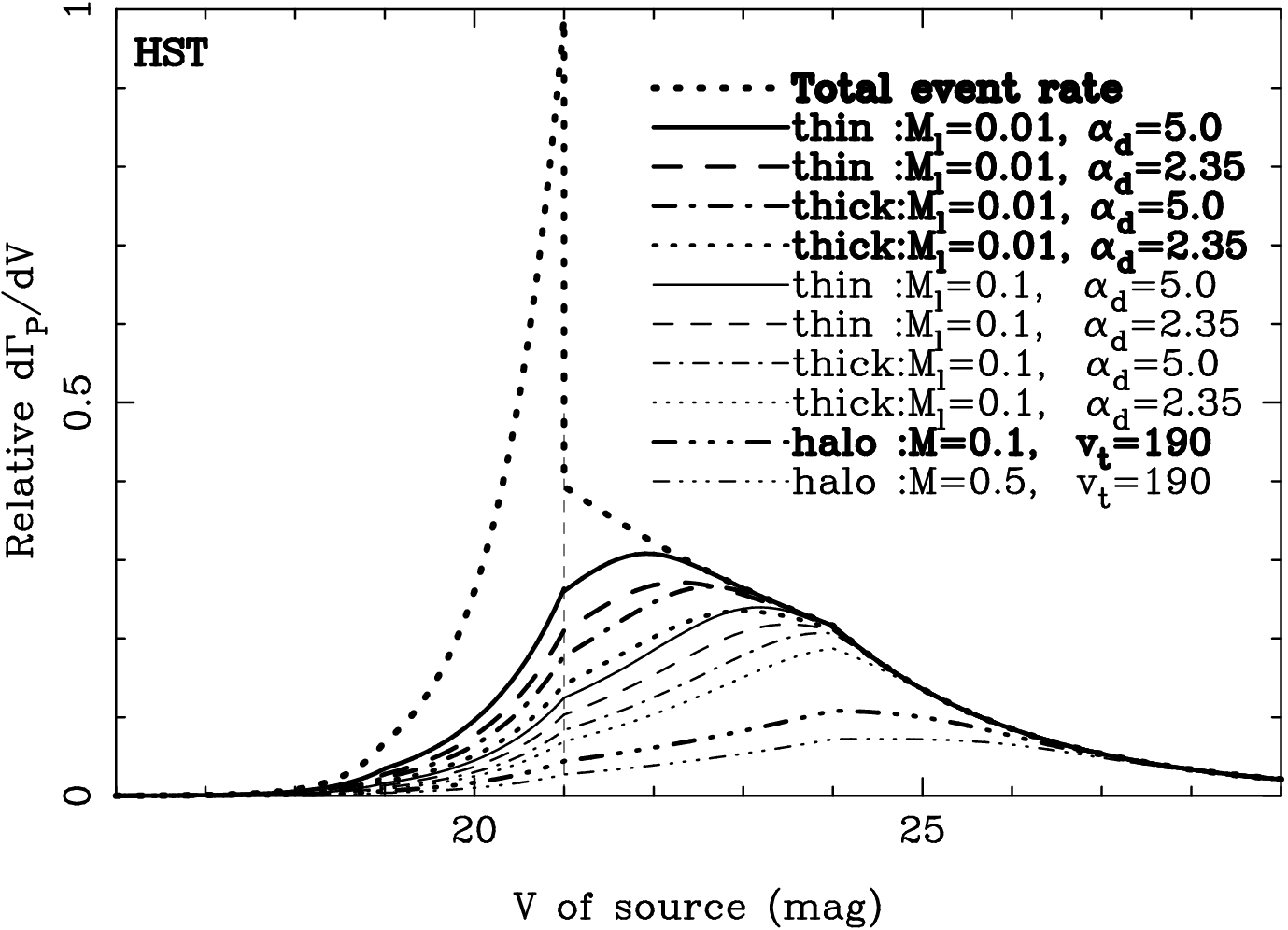}
\includegraphics[angle=-90,scale=0.35,keepaspectratio]{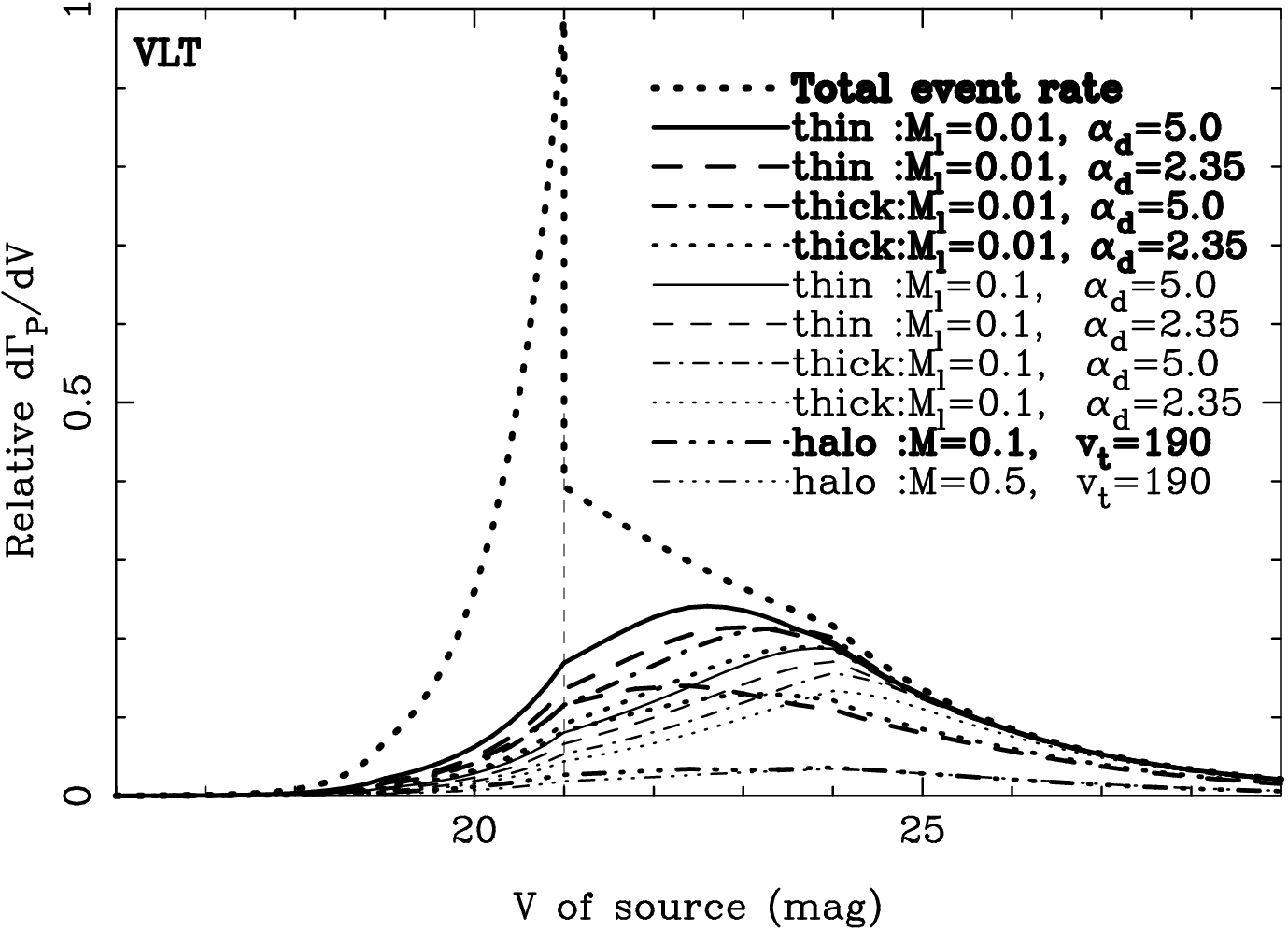}
\caption{Relative differential event rate distribution
  of parallax-measurable events $d\Gamma_{\!\rm P}/dV$ from the
  HST (upper panel) and from the VLT in the case of $\delta=45^\circ$ 
  (lower panel), normalized by $\Gamma_{\!\rm N}$ (for $V<V_{\rm obs}$) 
  or $\Gamma_{\!\rm E}$ (for $V>V_{\rm obs}$), as a function of source 
  magnitude $V$ and for various combinations of $M_{\rm l}$ and 
  $\alpha_{\rm d}$.
  By the VLT, for a comparison, we also show that distributions in the case of 
  $\delta=-367.5^\circ$ (i.e. the case that the observation start at $t_0$)
  for two cases of the thin and thick discs with $M_l = 0.01$ and 
  $\alpha_d=2.35$ as same line styles (lower one).
  We also show the relative $d\Gamma_{N}/dV$ and $d\Gamma_E/dV$ by the
  bold dotted line.
  \label{fig:ganmaP_hst}
  }
\end{center}
\end{figure}

In Fig. \ref{fig:ganmaP_hst} we show the relative event
rate distributions of parallax-measurable events $d\Gamma_{\!\rm P}/dV$
from the HST (upper panel) and from the VLT in the case of $\delta=45^\circ$
(lower panel), normalized by $\Gamma_{\!\rm N}$ (for $V<V_{\rm obs}$) or
$\Gamma_{\!\rm E}$ (for $V>V_{\rm obs}$) for thin disc, thick disc 
and halo events, as a function of source magnitude $V$ and for 
various combinations of $M_{\rm l}$ and $\alpha_{\rm d}$.  We
note that the absolute value of the vertical axis is not important
because these distributions are relative ones.  These
distributions are to be compared with the relative event rate
distribution for all events (the bold dotted line).  The left
side relative to the vertical dashed line corresponds to normal
microlensing events and the right side is for EAGLE events.
For disc events the greater $\alpha_{\rm d}$ or the smaller
$M_{\rm l}$, the higher $d\Gamma_{\!\rm P}/dV$ is. From Fig.
\ref{fig:ganmaP_hst} we can see that the event rate of
parallax-measurable EAGLE events $(V>21)$ is much higher than that of
normal microlensing events $(V<21)$.  We can also see that for halo events
the fraction of parallax-measurable events is still quite low even in EAGLE
events.
From the VLT (lower panel), we also show that distributions
in the case of $\delta=-367.5^\circ$ (i.e. the case that the 
observation starts at $t_0$) for two cases of the thin and thick discs
with $M_l = 0.01$ and $\alpha_d=2.35$ as same line styles (lower one).
From this figure, we can see that follow up observations by the VLT 
should be started $\sim 1$ day before $t_0$ to detect the parallax 
efficiently. Hereafter,  we assume $\delta=45^\circ$ for the case 
using the VLT.

We evaluate the ratio of the parallax-measurable EAGLE event rate to
the parallax-measurable normal event rate $\Gamma_{\!\rm P}(V>V_{\rm
  obs})/\Gamma_{\!\rm P}(V<V_{\rm obs})$ in Table \ref{tbl:endisc} for
disc events with $\Sigma_{\rm thin,thick}=50$  
$M_{\odot}$\,pc$^{-2}$ and in Table \ref{tbl:enhalo} for halo events 
with the lens mass of $M = 0.1$ and $0.5$ $M_{\odot}$ in the case of 
$V<25$ mag.  From Table
\ref{tbl:endisc} and \ref{tbl:enhalo}, it is clear that EAGLE events
enlarge the opportunity of parallax measurements.  The number of
parallax-measurable EAGLE events is $5 \sim 9$ (from the HST)
and $6 \sim 10$ (from the VLT) times larger than that of normal events
for disc events.  For halo events these are  $ \sim 8$ (from the HST)
and $4 \sim 6$ (from the VLT) times larger than that 
of normal events. 
We hereafter refer $\Gamma_{\!\rm P}$ as the parallax-measurable
EAGLE event rate because most of the parallax-measurable events are 
EAGLE events.

\begin{table}
 \centering

  \caption{Ratio of the parallax-measurable EAGLE event rate to
    the parallax-measurable normal event rate for disc events. 
    \label{tbl:endisc}}
  \begin{tabular}{@{}c@{}c@{\,}cc@{\,}c}\\
    $\alpha_{\rm d}$ & \multicolumn{2}{c}{the thin disc}
    & \multicolumn{2}{c}{the thick disc}
    \\
    & $M_{\rm l}=0.1 M_{\odot}$ & $M_{\rm l}=0.01 M_{\odot}$
    & $M_{\rm l}=0.1 M_{\odot}$ & $M_{\rm l}=0.01 M_{\odot}$ \\
  \hline
  \multicolumn{5}{c}{HST}\\
  \hline
  2.35 & 7.30 & 4.70 & 8.56 & 6.03\\
  5.0  & 6.71 & 4.05 & 8.07 & 5.36\\
  \hline
  \multicolumn{5}{c}{VLT}\\
  \hline
   2.35 & 8.44 & 5.78 & 9.50 & 7.24\\
   5.0  & 7.90 & 5.11 & 9.22 & 6.59\\
\end{tabular}
\end{table}

\begin{table}
 \centering
 \caption{Ratio of parallax-measurable event
    in EAGLE events to that in normal events for halo events.
    \label{tbl:enhalo}}
    \begin{tabular}{cccccc}\\
    \multicolumn{2}{c}{$v_{\rm t}$=190km s$^{-1}$}&
    \multicolumn{2}{c}{$v_{\rm t}$=220km s$^{-1}$} \\
    $M=0.1M_{\odot}$ & $M=0.5M_{\odot}$&
    $M=0.1M_{\odot}$ & $M=0.5M_{\odot}$ \\
  \hline
  \multicolumn{4}{c}{HST}\\
  \hline
  7.65 & 8.29 & 7.42 & 8.01\\
  \hline
  \multicolumn{4}{c}{VLT}\\
  \hline
  4.58 & 6.17 & 4.20 & 5.76\\
\end{tabular}
\end{table}

In Table \ref{tbl:ratio_disc} we show the fractions of
parallax-measurable EAGLE events out of all EAGLE events
$\Gamma_{\!\rm P}/\Gamma_{\!\rm E}$ and corresponding mean event time-scales
$\langle t_{\rm E} \rangle$ for disc events in the case of $V<25$ mag. For
$\Sigma_{\rm thick}(\Sigma_{\rm thin})=
30(70)$, $50(50)$ and $70(30)$ $M_{\odot}$\,pc$^{-2}$, 
the optical depths are $3.84(2.43)$, $6.41(1.74)$ and $8.97(1.04)$ 
in $10^{-8}$, respectively, if $f=1$, although the common factor $f$ 
is not essential so long as a relative event
rate is discussed, as noted in \S\,\ref{sec:model}. From Table
\ref{tbl:ratio_disc}, we see that $\langle t_E \rangle$ is strongly
affected by $M_{\rm l}$. $M_{\rm l} = 0.01 M_{\odot}$ seems to be
consistent with the observed value $t_{\rm E} \sim 40$ days
(\citealt{alc00b}) for the thick disc with $\alpha_{\rm d} = 2.35$.
$\Gamma_{\!\rm P}/\Gamma_{\!\rm E}$ only depends weakly on
$\Sigma_{\rm thin|thick}$ and $\alpha_{\rm d}$\footnote{
In our disc model, $v_{\rm t}$ is larger for the larger 
$\Sigma_{\rm thick}$ case, and so $\Gamma_{\!\rm P}/\Gamma_{\!\rm E}$ 
tends to be smaller. However for the thick disc from the HST it is 
inverse,  because $\Gamma_{\!\rm P} \propto$  $v_{\rm t}$, which  
is large for small $x$ and this effect 
is slightly larger than the former effect in this case. 
Anyway, these effects are 
quite small relative to the others we are concerned.
}.
We can see that the parallax effect can be measured in
$\sim75\%$ (from the HST) and $\sim 60\%$ (from the VLT)
of EAGLE events if the lenses are stars in the thick disc.

The fractions for the thin disc are $\sim10\%$ larger than that for
the thick disc, though the contribution of these components is small
because the optical depth is about $25\%$ relative to that of the
thick disc (\citealt{gou94a};  \citealt{gmb94}).  
$M_{\rm l}$ might be between $0.1$ and
$0.01$ from comparing the event durations with the observed value 
$t_{\rm E} \sim 40$ days.  Then, the parallax effect can also be measured in
$\sim75\%$ (from the HST) and $\sim 60\%$ (from the VLT)
of EAGLE events for the thin disc events.
In Table \ref{tbl:ratio_disc}, 
we also show that fractions in the case of $\delta = -367.5^\circ$ and 
$\Sigma_{\rm thick, thin}=  50$ $M_{\odot}$\,pc$^{-2}$ 
from the VLT. In this case the fractions are $\sim 30 \%$ smaller than 
that in the case of $\delta = 45^\circ$.

\begin{table*}
\begin{center}
  \caption{Fraction of parallax-measurable
    events in all EAGLE events and mean time-scales for disc events. }
  \label{tbl:ratio_disc}
  \begin{tabular}{@{}cccccccccc} \\
    $\alpha_{\rm d}$ & $\Sigma_{\rm thick}$ &
    \multicolumn{4}{c}{the thin disc} & \multicolumn{4}{c}{the thick disc}
    \\
     & ($M_{\odot}$ pc$^{-2}$) &
    \multicolumn{2}{c}{$M_{\rm l}=0.1M_{\odot}$}&
    \multicolumn{2}{c}{$M_{\rm l}=0.01M_{\odot}$}&
    \multicolumn{2}{c}{$M_{\rm l}=0.1M_{\odot}$}&
    \multicolumn{2}{c}{$M_{\rm l}=0.01M_{\odot}$}
    \\
    &  & $\langle t_E \rangle$ & $\Gamma_{\!\rm P}/\Gamma_{\!\rm E}$
    & $\langle t_E \rangle$ & $\Gamma_{\!\rm P}/\Gamma_{\!\rm E}$ 
    & $\langle t_E \rangle$ & $\Gamma_{\!\rm P}/\Gamma_{\!\rm E}$ 
    & $\langle t_E \rangle$ & $\Gamma_{\!\rm P}/\Gamma_{\!\rm E}$ \\
  \hline
  \multicolumn{10}{c}{HST}\\
  \hline
  2.35 &  30 & 61.7 & 0.659 & 23.2 & 0.867 & 128.9 & 0.513 & 48.6  & 0.745   \\
  2.35 &  50 & 55.6 & 0.656 & 21.0 & 0.865 & 118.8 & 0.515 & 44.8  & 0.747   \\
  2.35 &  70 & 49.7 & 0.654 & 18.7 & 0.863 & 109.6 & 0.517 & 41.3  & 0.749   \\
   \\
  5.0  &  30 & 41.3 & 0.732 & 13.0 & 0.938 & 86.3 & 0.586 & 27.2  & 0.832   \\
  5.0  &  50 & 37.2 & 0.729 & 11.7 & 0.936 & 79.5 & 0.588 & 25.0  & 0.834   \\
  5.0  &  70 & 33.3 & 0.727 & 10.5 & 0.934 & 73.4 & 0.590 & 23.1  & 0.835   \\
  \hline
  \multicolumn{10}{c}{VLT ($\delta = 45^\circ$)}\\
  \hline
  2.35 &  30 & 61.7 & 0.486 & 23.2 & 0.704 & 128.9 & 0.355 & 48.6  & 0.580   \\
  2.35 &  50 & 55.6 & 0.479 & 21.0 & 0.687 & 118.8 & 0.356 & 44.8  & 0.575   \\
  2.35 &  70 & 49.7 & 0.469 & 18.7 & 0.664 & 109.6 & 0.355 & 41.3  & 0.566   \\
   \\
  5.0  &  30 & 41.3 & 0.555 & 13.0 & 0.785 & 86.3 & 0.421 & 27.2  & 0.668   \\
  5.0  &  50 & 37.2 & 0.546 & 11.7 & 0.765 & 79.5 & 0.420 & 25.0  & 0.660   \\
  5.0  &  70 & 33.3 & 0.532 & 10.5 & 0.736 & 73.4 & 0.418 & 23.1  & 0.648   \\
  \hline
  \multicolumn{10}{c}{VLT ($\delta = -367.5^\circ$)}\\
  \hline
  2.35 &  50 & 55.6 & 0.359 & 21.0 & 0.451 & 118.8 & 0.291 & 44.8  & 0.420   \\
  5.0  &  50 & 37.2 & 0.393 & 11.7 & 0.484 & 79.5 & 0.336 & 25.0  & 0.465   \\
\end{tabular}
\end{center}
{\flushleft Note: The event time-scales $\langle t_{\rm E} \rangle$ are given in day.}
\end{table*}

For halo events, we adopt the mass function as a delta function 
$\delta(M)$, a Gaussian $G(M)$  and a log-normal Gaussian $G(logM)$  
distributions because the lens mass function is not well known. For $\delta(M)$ 
we adopt $M= 0.1$ and $0.5 M_{\odot}$. For $G(M)$ and $G(logM)$  
we take the mean of the mass
$M=0.1$ and 0.5 $M_{\odot}$. The variance is
$0.4M_{\odot}$ for the  $G(M)$ and
$\log(0.4M_{\odot}/M)$ for  $G(logM)$. Of
course the mass function is zero for negative $M$ in $G(M)$. 
The typical transverse velocity is $v_{\rm t}= 190$
or $220$ km\,s$^{-1}$. The estimated fractions 
in the case of $V<25$ mag are shown 
in Table \ref{tbl:ratio_halo}.
The optical depth is $4.8 \times 10^{-7}$ if $f=1$.  From Table
\ref{tbl:ratio_halo}, we can see that the fraction of parallax-measurable
events out of all EAGLE events is $\sim 20\%$ (from the HST) and 
$\sim 10\%$ (from the VLT).

The fraction of parallax-measurable events depends on the lens location.
Then we can also statistically discriminate whether the lenses are in 
the thick disc or halo, by using
the parallax-measurable EAGLE event rate.

\begin{table}
\begin{center}
 \caption{Fraction of parallax-measurable events in all EAGLE
          events and mean time-scales for halo events with mass functions, 
          $\delta(M)$: the delta-function with mass $M$,
           $G(M)$: the Gaussian and $G(logM)$: the
          log-normal Gaussian with mean mass $M$. 
          }
 \label{tbl:ratio_halo}
   \begin{tabular}{cccccccccc}\\
    Mass & $M$ 
    & \multicolumn{2}{c}{$v_{\rm t}=190$km\,s$^{-1}$}
    & \multicolumn{2}{c}{$v_{\rm t}=220$km\,s$^{-1}$} \\
    Func.&($M_{\odot}$) 
    & $\langle t_{\rm E} \rangle$ & $\Gamma_{\!\rm P}/\Gamma_{\!\rm E}$ 
    & $\langle t_{\rm E} \rangle$ & $\Gamma_{\!\rm P}/\Gamma_{\!\rm E}$ \\
  \hline
  \multicolumn{6}{c}{HST}\\
  \hline
  $\delta(M)$ & 0.1 & 22.5 &  0.306 &  19.4 & 0.289  \\
              & 0.5 & 50.3 &  0.201 &  43.4 & 0.191  \\
  \\
  $G(M)$      &0.1 &  46.0 &  0.221 &  39.8 & 0.210  \\
              &0.5 &  57.1 &  0.194 &  49.3 & 0.184  \\
  \\
  $G(logM)$   &0.1 &  46.0 &  0.251 &  39.8 & 0.238  \\
              &0.5 &  51.2 &  0.200 &  44.2 & 0.190  \\
  \hline
  \multicolumn{6}{c}{VLT}\\
  \hline
  $\delta(M)$ & 0.1 & 22.5 &  0.122 &  19.4 & 0.104  \\
              & 0.5 & 50.3 &  0.105 &  43.4 & 0.095  \\
  \\
  $G(M)$      &0.1 &  46.0 &  0.108 &  39.8 & 0.096  \\
              &0.5 &  57.1 &  0.100 &  49.3 & 0.091  \\
  \\
  $G(logM)$   &0.1 &  46.0 &  0.108 &  39.8 & 0.095  \\
              &0.5 &  51.2 &  0.104 &  44.2 & 0.094  \\
\end{tabular}
\end{center}
Note: The event time-scales $\langle t_{\rm E} \rangle$ are given in day.
\end{table}

\section{Discussion and Conclusion}
\label{sec:disc}

We have seen that the EAGLE event rate is as high as that for normal
events even for disc events towards the LMC. Since the period in which
sources are visible ($V <21$) is usually short (1 day $\sim$ 40 days),
the detection efficiency heavily depends on the observational frequency.
The observational programs currently undertaken by most groups are not
adequate. Hourly monitoring with a 1-m class dedicated telescope and
the real-time detection of EAGLE events by the DIA are required to 
issue alerts with a high detection efficiency.
However, for the events in which the source star is $V>25$ this period is
less than 2 days, so it is  
difficult to detect. Thus we estimated $\Gamma_{\!\rm E}/\Gamma_{\!\rm N}$
and $\Gamma_{\!\rm P}/\Gamma_{\!\rm E}$ in the case that
the source star is $V<25$.
These are decreased but still sufficiently high.
Of course, the larger alert telescopes make it easier and faster to issue
the alerts.

Estimating the parallax-measurable event rate, we advocate follow-up 
observations with a space telescope such as the HST and an 8-m class 
ground-based telescope such as the VLT.
We have found that EAGLE events enlarge the opportunity of parallax 
measurements by 
$5 \sim 9$ (from the HST) and $6 \sim 10$ (from the VLT) 
times for disc events, and by
$\sim 8$ (from the HST) and $4 \sim 6$ (from the VLT) times 
for halo events relative to that in normal microlensing events.  
We have also found we can measure the parallax
effect in $\sim75\%$ (from the HST) and $\sim 60\%$ (from the VLT)
of EAGLE events if the lenses are in the thick or thin disc, and in
 $\sim 20\%$ (from the HST) and $\sim 10\%$ (from the VLT)
if the lenses are in the halo. 
Since $\tilde{v}$ for halo objects are $5 \sim 8$ times
larger than that for disc stars, we can determine whether the
lens objects are in the halo or discs for each event.
We can also statistically constrain the lens locations
by using the parallax-measurable EAGLE event rate.

In follow-up observations from the HST, in this paper we assumed that
the observations start just after the peak as the most conservative case.
Of course, an observation around the peak is better than that just after 
the peak.
However predicting $t_0$ is not so easy for very faint source events. 
Extensive follow-up observations from small ground-based telescopes 
around the world are needed to predict $t_0$ and inform to the HST.
Furthermore a flexible operating program of the HST for the alerts are 
required. If the alert telescope is at a latitude $-30^\circ$, the alert
can be issued for half of the year (Southern Summer),
while at a latitude $-44^\circ$ (such as New Zealand) it can be done 
all year round.
The total operation time of the HST would be several days per year.

From the VLT, we assumed $\delta = 45^\circ$.
However for the events with faint source ($V>24$) the time until
$t_0$ is short, and the beginning of the follow-up would tend to be delayed.
A delay of $\sim 1$ day reduces the possibility of measuring the parallax
by $\sim 30 \%$ as shown in \S\,\ref{sec:results}.
This observation can be done for only half of the year (Southern Summer).


The true source flux $f_0$ is needed to measure the precise value of
$\tilde{v}$ in the light curve fitting.
Then follow-up observations by a high resolution telescope such as the HST
are needed to get an accurate $f_0$ after the event.

In short, a practical observation strategy would be
to observe hourly with a 1-m class telescope and perform
real-time analysis with DIA to issue alerts
to world observatories and the HST or 
the VLT for follow-up observations.
Then after the events  $f_0$ should be measured by the HST.


To demonstrate this specifically we estimate the number of expected
parallax-measurable events for the two cases that these are mainly halo
events or disc events.  In both cases, the thin disc events
are included.  For the typical
parameters $\alpha_{\rm d} = 2.35$, $\Sigma_{\rm thin}=\Sigma_{\rm thick}=50 
M_{\odot}$pc$^{-2}$, $V_{\rm th} = 20$,
a detection efficiency of 50\% and source stars of $V<25$,
one can expect to find $\simeq$13 EAGLE
events from 3-year observations of 11 square degrees of the LMC
central region (as the MACHO collaboration does).
In the case of follow-up from the VLT, 
the expected number would be half.  In these 13 EAGLE
events, $\sim 2$ events $(15 \% )$ are due to the stars in the thin disc and
a further $\sim 11$ events are due to MACHOs or the dark stars in the 
thick disc.
In considering these 11 events, reasonable parameters are $M = 0.1$
or $0.5 M_{\odot}$ except $M = 0.1M_{\odot}$ in the case of the
$\delta$-function for halo events, and $M_{\rm l} = 0.01M_{\odot}$ for
disc events, to be consistent with $\langle t_E \rangle\simeq$40 days
(\citealt{alc00b}).  In this case, we will be able to measure $\tilde{v}$
in $\sim 10$ (from the HST) or $\sim 4$ (from the VLT) events
for disc events, $\sim 4$ (from the HST) or $\sim 1$ (from the VLT) event
for halo events, which include thin disc events. 
We can constrain lens locations  strongly based on the 3-year 
statistics of these observations.

In conclusion, one could statistically discriminate whether the typical lens 
locations are in a thick disc or not, using parallax measurements even with 
$\sim R_\oplus$ scale baseline.
One could also distinguish whether the
lenses are MACHOs or stars in the LMC itself through finite source 
size effects measurements in EAGLE events (\citealt{sum00}). Therefore
one could identify lens objects as halo MACHOs, dark stars in the
Galactic thick disc, or stars in the LMC through these observations.

A real-time alert system with DIA, has been
introduced by the MOA collaboration\footnote{see {\tt
http://www.phys.canterbury.ac.nz/\~{}physib/alert/alert.html}} from 2000
and by the OGLE collaboration\footnote{see {\tt
http://www.astrouw.edu.pl/\~{}ogle/ogle3/ews/ews.html}} from 2002.

\section*{Acknowledgments}
We are grateful to Y.~Muraki for his supervision, and also to P.~Yock 
and I.~Bond for helpful comments. We would like to thank K. Sahu, 
B. Paczy\'{n}ski, L. Eyer and J. Tan for fruitful comments. 
We also acknowledge helpful discussions and continuous encouragement to
Y.~Sofue, S.M.~Miyama, N.~Sugiyama, and R.~Nishi.
T.S. acknowledge the financial support from the Nishina Memorial Foundation.
We are also thankful to the referee for suggestive comments.

\label{lastpage}
\clearpage

\end{document}